\documentclass[a4paper,twocolumn]{esapub} 
\usepackage{times, epsfig}
\title{IBIS/PICsIT Status}
\author[1]{G. Di Cocco}
\author[1]{V. Bianchin}
\author[1]{L. Foschini}
\author[1]{F. Gianotti}
\affil[1]{INAF/IASF-Bologna, via Gobetti 101, 40129 Bologna, Italy (dicocco, bianchin @iasfbo.inaf.it)}
\author[2]{P. Laurent}
\affil[2]{Service d'Astrophysique -- SAp-CEA, Saclay, France}
\author[1]{G. Malaguti}
\author[3]{L. Natalucci}
\affil[3]{INAF/IASF-Roma, Via Fosso del Cavaliere, 00100 Roma, Italy}
\author[1]{F. Schiavone}

\begin{document}
\maketitle
\begin{abstract}
We report about the status of the PICsIT layer of the imager IBIS. 
The instrument has been tested in both Spectral Imaging and Spectral Timing
Mode. 
PICsIT Single Events, Multiple and ISGRI data of the Crab are simultaneously
fitted to a power law model and fit parameters are compared with the standard
values. 
We present a joint fit of SPI, ISGRI, PICsIT data of the long GRB041219, that
was seen by the PICsIT instrument in both Spectral Imaging and Spectral Timing
acquisition modes. This allowed us to generate a preliminary version of the
instrumental response matrices for the PICsIT Spectral Timing mode.

\end{abstract}
\section{PICsIT Outline}
PICsIT (Pixellated Imaging CsI Telescope) is the high-energy detector ($175 ~
\rm{keV}$-$13 ~ \rm{MeV}$) of the imager IBIS on board of the INTEGRAL
satellite. 
The instrument provides fine imaging, good sensitivity and moderate energy
resolution over a wide energy interval in the hard X/gamma-ray domain. 
PICsIT is composed by $4096 ~ (64\times 64)$ CsI(Tl) crystals organised in 16
semi-modules, with independent electronic units. 
The imager collects the shadowgram cast by the IBIS coded mask, covering a
FOV=$29^{\circ} \times 29^{\circ}$ (zero response) sampled in $10'$ (see
\cite{dcc} for more details).  

Data are on-board equalised and pre-processed by means of the LUT (Look Up
Tables) according to the event type and the detector operational mode. 
A valid PICsIT event is every signal that is not in coincidence with ISGRI
events (Compton events), Calibration events or VETO strobes. 
PICsIT detects and stores separately two types of event: Single and Multiple
Events. 
Single Events are made of detections in one single pixel, produced by
energy deposits above the detector low threshold. 
The energy range for single events is $175~\rm{keV} - 6.5~\rm{MeV}$. 
Multiple (Double) Events are generated by a photon energy deposit of $350
~ \rm{keV} - 13~ \rm{MeV}$ across two or more pixels. 
Due to the electronic separation of semi-modules Multiple Events across
semi-modules are treated as two separate Single Events (see \cite{bbb}). 

Data acquisition and preprocessing follow two operational modes:
Photon-by-Photon Mode (PPM) and Standard Mode (SM). 
In PPM the instrument reveals the complete photon information: event position,
time delay between consecutive events and energy in $1024$ channels for both
Singles and Multiples. 
Because of the satellite telemetry restrictions the acquisition in PPM is
activated only for Calibration purposes and during Slews ($\sim 120 ~
\rm{s}$). 
Moreover in PPM the energy of the transmitted data is limited to $\sim 1 ~ 
\rm{MeV}$, again because of tight telemetry budget. 
During pointings, PICsIT is switched in its nominal operational mode, the SM,
which is composed by two complementary Submodes: Spectral Imaging (SI) and
Spectral Timing (ST). 
In SI the full spatial information on $64 \times 64$ pixels is kept but events
are on-board integrated over $\sim 2 ~\rm{ks}$ and stored in 256 energy
channels. 
Conversely, ST mode allows a time binning from $0.97 ~ \rm{ms}$ to $500 ~
\rm{ms}$, up to $8$ energy bins, but no spatial resolution. 
From Revolution $441$ (2006 May 24) the conversion table for ST (containing
Singles and Multiples together) has been updated. 
Time binning is $16 ~ \rm{ms}$ in $8$ energy bands: $208-260 ~ \rm{keV}$,
$260-312 ~ \rm{keV}$, $312-364 ~ \rm{keV}$, $364-468 ~ \rm{keV}$, $468-572 ~
\rm{keV}$, $572-780 ~ \rm{keV}$, $780-1196 ~ \rm{keV}$, $1196-2600 ~
\rm{keV}$. 

More information on the status of the detector can be found at the IBIS/PICsIT
web page in
Bologna\footnote{\texttt{http://www.iasfbo.inaf.it/index.php?option=}  
\hspace{0.5cm} 
\texttt{com\_content\&task=view\&id=42\&Itemid=49}}

\section{The Crab spectrum in Spectral Imaging Mode}
\begin{figure}[h!]
\centering
\includegraphics[angle=-90,scale=0.35]{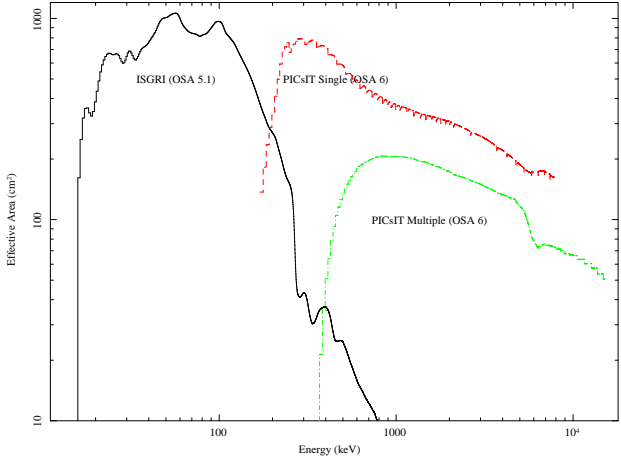}
\caption{Effective area for IBIS instruments. For ISGRI we use the spectral
  response of \texttt{OSA 5.1}; PICsIT effective areas for Single and Multiple
  Events are the updated versions in \texttt{OSA 6}. \label{fig1}} 
\end{figure}

\begin{figure}[h!]
\centering
\includegraphics[angle=-90,width=1.0\linewidth]{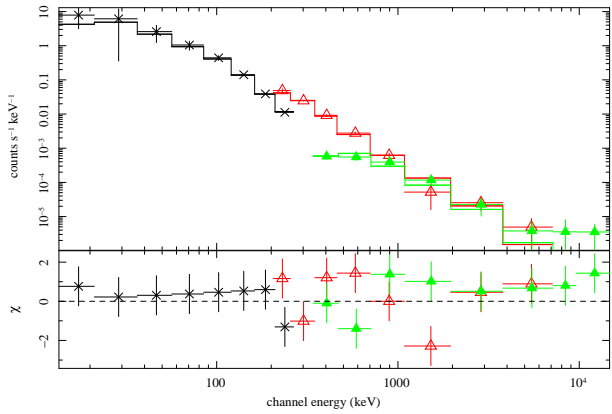}
\caption{Crab spectrum: joint fit of ISGRI (crosses), PICsIT singles (open
  triangles) and PICsIT multiples (filled triangles) data. Residuals are
  plotted in $\sigma$ units. \label{fig2}} 
\end{figure}

We tested the spectral capabilities of the IBIS/PICsIT layer on a joint fit of
PICsIT and ISGRI data of the Crab. 
Spectral extraction can follow two methods: the first one is to run up to
level SPE the pipeline of the Offline Scientific Analysis
(\texttt{OSA})\footnote{\texttt{http://isdc.unige.ch/?Soft+download}}, 
whose algorithms for IBIS are described in \cite{gdf}.  
However, because of the generally low signal/background ratio  of sources
in PICsIT operational energy range, this procedure works well with the
integration of shadowgrams before spectral extraction. 
This is possible only in staring mode acquisition, while during scientific
pointings the satellite usually takes a set of $\sim 2 ~ \rm{ks}$ acquisitions
(Science Windows - ScWs) following a dither pattern. 
Improvements on this method are under development, but to date the most stable
procedure is spectral extraction through the imaging pipeline. 
The scientific analysis software is run up to level IMA, when shadowgrams are
deconvolved and images are obtained for each ScW. 
Subsequently the total mosaic of the whole data set is obtained by the
\texttt{varmosaic} tool in \texttt{HEASoft v.6.1.1}. 
Counts and errors in each energy bin are then derived from the intensity and
significance maps. 

It is important to remind that the low Signal to Noise Ratio (SNR) requires
long exposure time. 
For this reason our data set covers all the Crab observations available up to
date, from February 2003 to May 2005 (Revolutions 39, 40, 41, 42, 45, 170,
239, 300, 365 and 422). 
Revolutions 102 and 103 have been neglected since the detector underwent high 
temperature fluctuations ($\sim 10^{\circ}$) that strongly affected imaging
performances. 
Data have been downloaded from the ISDC data archive\footnote{\texttt{http://isdc.unige.ch/index.cgi?Data+browse}} and we selected
Observations in which PICsIT was in Standard operational Mode and where both
Single and Multiple Events were collected. 
Data sample has been cleaned from ScWs where problems
of different nature have been recognised: ScWs with exposure lower than $1
\rm{ks}$, with missing calibration files, with badly downloaded histograms and
so on. 

PICsIT data analysis was carried out by means of \texttt{OSA 6}, running the
pipeline on the 8 standard energy bins for single and multiple events \cite{valrep}. 
The total mosaic of data set has been produced, with an effective
exposure of $1.1 ~ \rm{Ms}$. 
The long total exposure enhances the SNR and allows a full energy range
investigation of the source spectrum. 
We included in the spectral analysis also data with very low values of the SNR
since the Crab is a known source and we are mainly interested in analysing fit
problems. 
For Single and Multiple events we used the newly released spectral responses
(RMF/ARF), included in \texttt{OSA 6} (see Fig.\ref{fig1}). 

\begin{figure*}[ht!]
\centering
\includegraphics[scale=0.4]{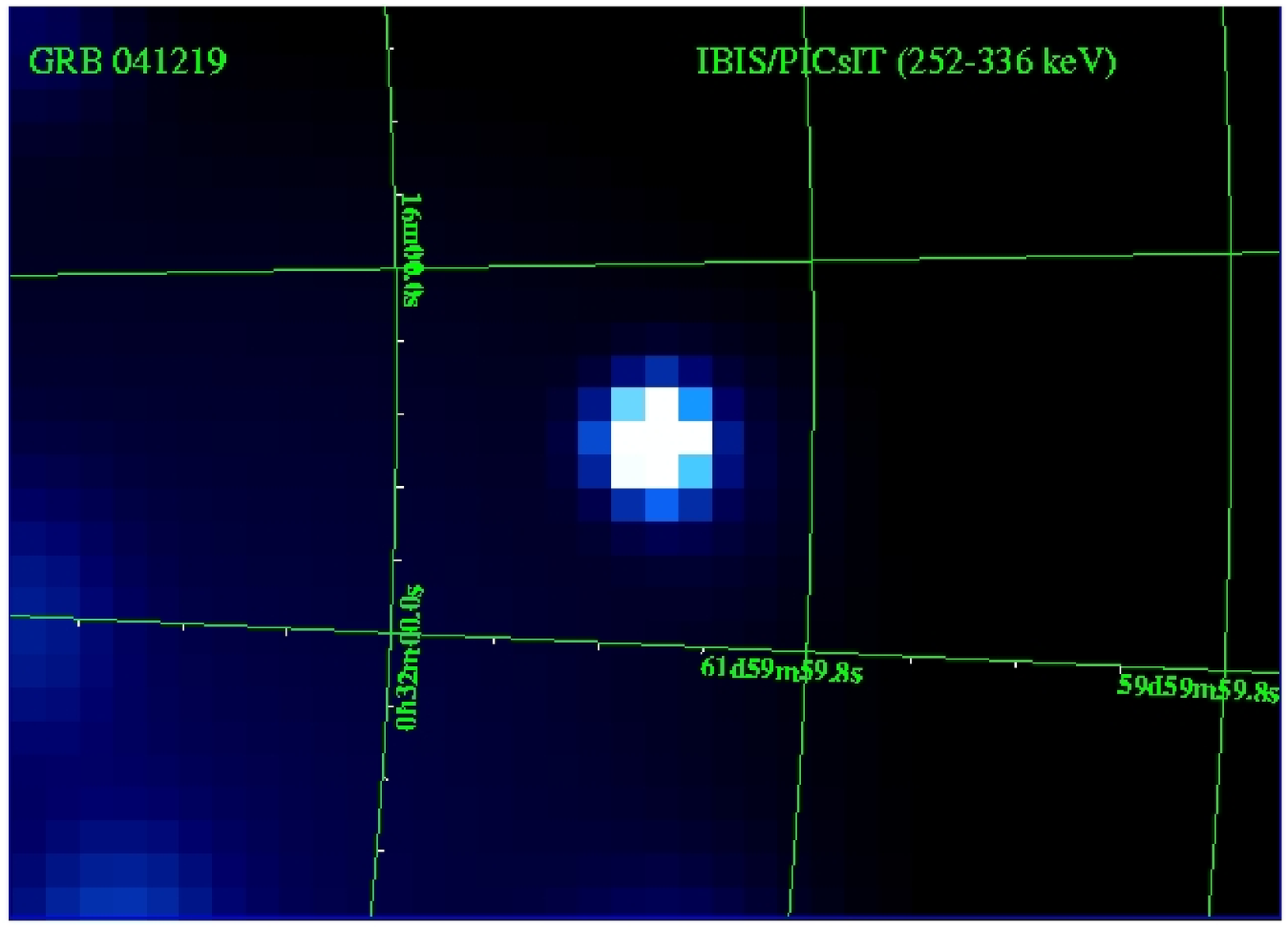}
\includegraphics[scale=0.28]{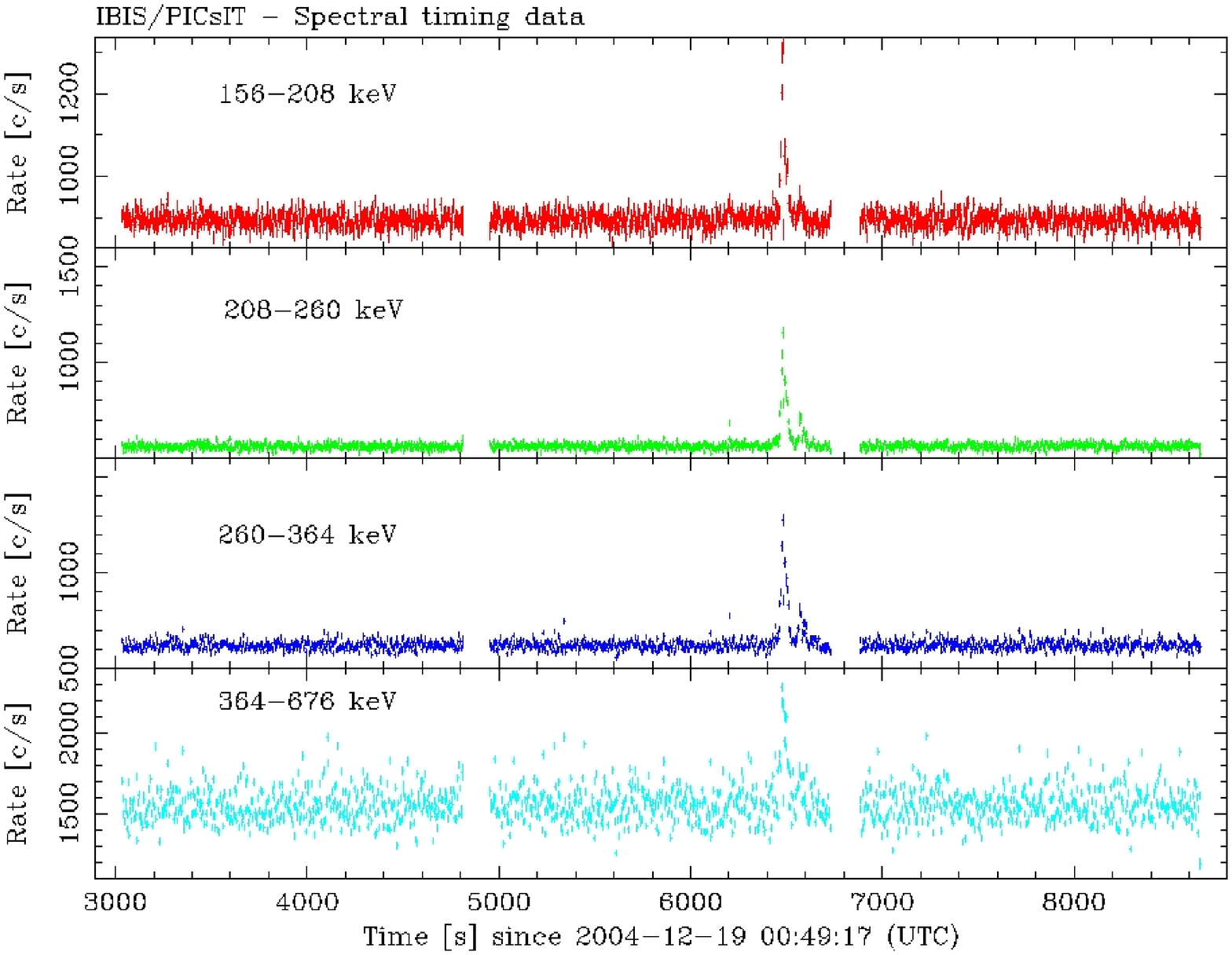}
\caption{Image ($252-336 ~ ~ \rm{keV}$) and lightcurves of the very long GRB041219, detected in both Spectral Imaging (left panel) and Spectral Timing (right panel). \label{fig3}}
\end{figure*}
From the same data set we extracted the ISGRI spectrum running the \texttt{OSA
  5.1} pipeline up to level SPE. 
We used \texttt{XSPEC v.11.3.2} to perform a joint fit of PICsIT Single,
PICsIT Multiple and ISGRI data using a standard unabsorbed power law model
(Fig.\ref{fig2}). 
A systematic error of $1\%$ is added to ISGRI data and a $5\%$ systematic
error is considered only in the first bin for PICsIT Single Events, which
is strongly affected by cosmic rays induced events \cite{slb}.
The best-fit model ($\chi^2 = 23.8$ for $20$ dof, Prob.$=0.25$) is a power-law
with photon index $\Gamma = 2.11 ^{+0.09}_{-0.08}$ and normalisation at
$1$~keV $N=8.1 ^{+4.9}_{-2.8}$ ph~cm$^{-2}$~s$^{-1}$~keV$^{-1}$. 
We obtain a good accordance with the ``standard'' values $\Gamma = 2.10 \pm 0.03$
and $N=9.7 \pm 1.0$ ph~cm$^{-2}$~s$^{-1}$~keV$^{-1}$ by \cite{ts}. A more
recent analysis by \cite{mk}, based on the observation of several satellites
in the $0.1-1000$~keV energy range, gave similar results ($\Gamma=2.08$,
$N=8.97$~ph~cm$^{-2}$~s$^{-1}$~keV$^{-1}$).
In the $20~ \rm{keV} - 10 ~ \rm{MeV}$ range the model flux is $\approx 0.26
^{+0.31}_{-0.14}$ ph cm$^{-2}$ s$^{-1}$, which is still in accordance with the
expected value $\approx 0.33 ^{+0.12}_{-0.10}$ ph cm$^{-2}$ s$^{-1}$ by
\cite{ts}, due to the large error affecting the normalization parameter.  
The intercalibration coefficients for PICsIT single events ($C_S =0.52
^{+0.04}_{-0.03}$) and for multiples ($C_M = 0.31 \pm 0.06$) are sensibly
low as compared with ISGRI (fixed to $1$ as reference parameter).
This problem requires a future deeper investigation on the response matrix for
both PICsIT single and multiple events.

\section{Calibration of Spectral Timing Mode}

\begin{figure}
\centering
\includegraphics[angle=-90,width=1.0\linewidth]{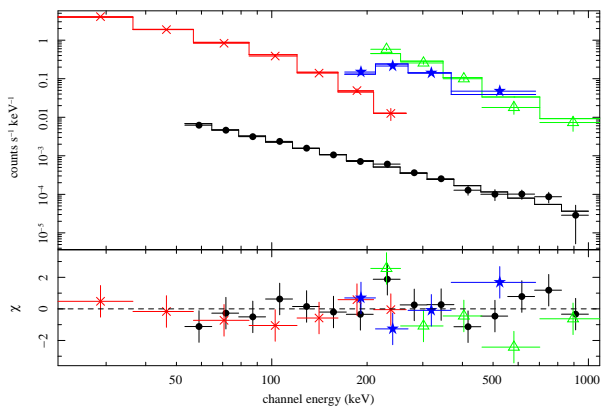}
\caption{GRB041219 spectrum: joint fit of SPI (circles), ISGRI (crosses),
  PICsIT Spectral Imaging (triangles) and PICsIT Spectral Timing (stars)
  data. Residuals are given in $\sigma$ units. \label{fig4}} 
\end{figure}
In Spectral Timing Mode, PICsIT single and multiple events are not individually
treated. 
Therefore data acquired in this mode can be straightforwardly used to build
lightcurves of the whole detector plane but spectral extraction needs a proper
response matrix. 
To date, no RMF and ARF are available and a source dependent calibration of the Spectral Timing Mode
could be attempted when there are simultaneous detections in both Spectral Timing and
Spectral Imaging Mode. This occurred during the long GRB041219 ($\sim 360$~s,
see Fig.~\ref{fig3}) and we used this observation to generate a preliminary
version of the spectral response for ST mode data.
We perform a simultaneous fit with ISGRI and SPI data in order to better
determine model parameters and to check the new ST matrix. 
Note that when the GRB occurred the ST was set to $4 ~ \rm{ms}$ time resolution
and 4 energy bands ($156-208$, $208-260$, $260-364$, $364-676 ~ \rm{keV}$). 

We extracted the spectrum of GRB041219 for ISGRI, SPI, PICsIT Single
Events acquired in Spectral Imaging and PICsIT Spectral Timing. 
For ISGRI and SPI we run the standard \texttt{OSA 5.1} pipeline up to level
SPE, while for PICsIT SI we followed the procedure described above, deriving
counts and errors from the image step. For PICsIT ST we extracted source plus 
background counts from the lightcurves of the ScW where the GRB occurred,
while the background to be subtracted was extracted by using the ScWs before and
after the event (see Fig.~\ref{fig3}, right panel).

Using \texttt{XSPEC} we fitted data to an unabsorbed power law (Fig.~\ref{fig4}). A first fit
provides a rather poor $\tilde{\chi}^2\approx 2$. Most of the problems are due
to different systematic effects (cosmic-rays induced events in the low energy
bins of PICsIT, the saturation of ISGRI during the GRB, other effects in SPI, etc...)
and, therefore, for this preliminary analysis, we decided to assign an overall 
value of $7\%$ for systematics. 

The best-fit model ($\chi^2 = 31.1$ for $26$ dof with Prob. $=0.23$) has
photon index $\Gamma = 1.89 \pm 0.05$ and normalisation at $1$~keV $N = 14.8
^{+4.2}_{-2.4}$~ph~cm$^{-2}$ s$^{-1}$~keV$^{-1}$. 
In this case the reference for the intercalibration constant is SPI, since
ISGRI  is saturated by the high level counts of the source. 
This is also shown in the low value of the intercalibration constant, that is 
$C_{ISGRI}=0.21 ^{+0.01}_{-0.02}$.
The best-fit intercalibration for PICsIT single events in SI is $C_{PSI} = 0.9
\pm 0.1$ and for PICsIT ST is $C_{PST} = 0.54 ^{+0.07}_{-0.04}$. 

We would like to underline that these are preliminary results and studies on
the instrumental response of PICsIT in ST mode are currently on going.

\section{Conclusions and Summary}
\begin{figure*}
\centering
\includegraphics[width=0.8\linewidth]{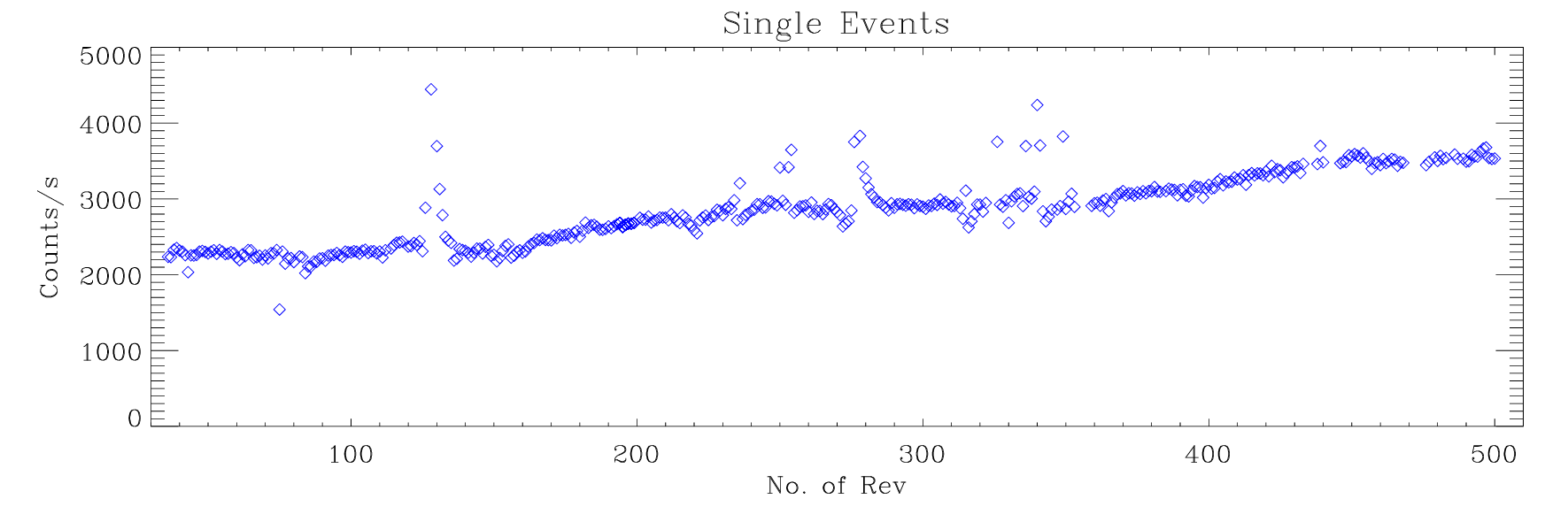}
\includegraphics[width=0.8\linewidth]{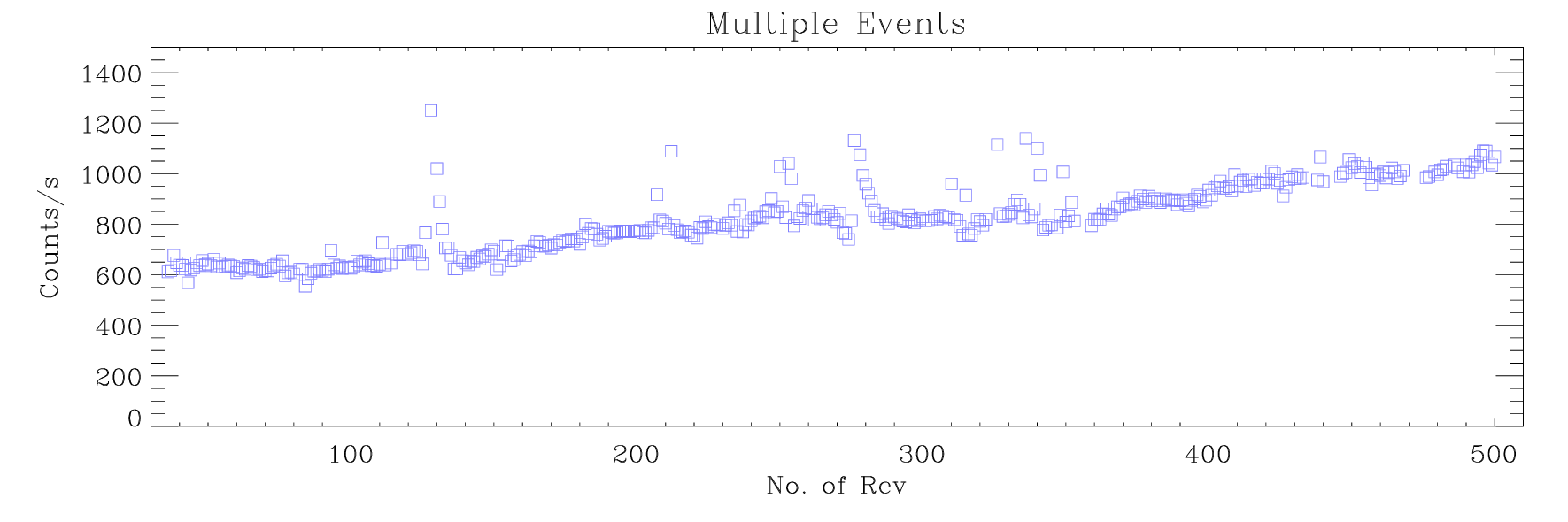}
\caption{PICsIT background level for single (upper panel) and
  multiple (lower panel) events, from Revolution 36 to 505 . \label{fig5}}
\end{figure*}

In this work we report the status of the IBIS/PICsIT instrument on board of the
INTEGRAL satellite. 
After the launchof the satellite, a contiunuous monitoring of PICsIT was
activated. 
Results obtained to date (e.g. Fig.~\ref{fig5}) confirm the functional
stability of instrument hardware. 
The number of dead pixels remained constant since the very beginning of
the mission, for a total of 52 killed pixels. 
The average gain is in agreement with the value of $7.1$ keV/ch, obtained
from in-flight calibrations (\cite{mbb}, \cite{dmf}). 
The background level shows a low but continuous 
increase, likely to be due to solar activity (see Fig.~\ref{fig5}). 

PICsIT spectral capabilities have been verified using all the available
observations of the Crab.  
The results of a joint fit of PICsIT and ISGRI data are in reasonably good
agreement with the expected values. 
We focused on the possibility of full range analysis with long exposure
data set and the need of future more detailed investigation of the
instrument response.
We present the first spectral analysis of PICsIT Spectral Timing data for long
GRB041219, that was detected in both acquisition modes. 

Up to date PICsIT has revealed four compact object: two pulsars (Crab and PSR
B1509-58) and two galactic black holes (Cygnus X-1 and XTE
J1550-564\footnote{A recent study of the 2003 outburst of this Galactic Black
  Hole, including PICsIT data, has been presented in \cite{vb}.}). 
Moreover, since the Spectral Timing mode is particularly effective in
detecting powerful impulsive emissions, the list of GRBs is continuously
updated and now contains 47 events. 
An updated PICsIT source catalogue can be found in the instrument web
page\footnote{\texttt{http://www.iasfbo.inaf.it/index.php?option=}
  \texttt{com\_content\&task=view\&id=42\&Itemid=49}}.

\end{document}